\documentclass[12pt]{article}
\usepackage{graphicx}
\usepackage{amsfonts}
\usepackage{amsmath,amssymb}
\usepackage{mathrsfs}
\usepackage{setspace}
\usepackage{cite}

\def\beq{\begin{equation}}
\def\eeq{\end{equation}}
\def\bea{\begin{eqnarray}}
\def\eea{\end{eqnarray}}
\def\no{\noindent}
\def\nn{\nonumber}

\def\ba{\begin{array}}
\def\ea{\end{array}}

\def\d{\partial}

\def\l{\langle}
\def\r{\rangle}
\def\v{\vert}

\begin{document}

\begin{center}
{\Large \bf \sf
      Exact solution of $D_N$ type quantum Calogero model \\
        through a mapping to free harmonic oscillators }

\vspace{1cm}
{\sf Pratyay Banerjee\footnote{e-mail: pratyay.banerjee@saha.ac.in} and 
B. Basu-Mallick \footnote{% Corresponding Author:
e-mail: bireswar.basumallick@saha.ac.in %, Phone: +91-33-2337-5346,
%FAX: +91-33-2337-4637
}}

\bigskip

{\em Theory Group, Saha Institute of Nuclear Physics, \\
1/AF Bidhan Nagar, Kolkata 700 064, India}

\end{center}

\bigskip

\vspace {1.4 cm}
\no{\bf Abstract }
\baselineskip=16pt
\vskip .2 cm 
We solve the eigenvalue problem of the $D_N$ type of Calogero model by mapping it to a set of 
decoupled quantum harmonic oscillators through a similarity transformation. 
In particular, we construct the eigenfunctions of this Calogero model  
from those of bosonic harmonic oscillators having  
either all even parity or all odd parity. It turns out that
the eigenfunctions of this model are orthogonal with respect
to a nontrivial inner product, which can be derived from the
quasi-Hermiticity property of the corresponding conserved quantities.

\vspace{.8 cm}
%\no PACS No.: 02.30.Ik, 75.10.Jm, 05.30.-d %, 03.65.Fd

\vspace {.8 cm}
%\no Keywords: Haldane-Shastry spin chain, vertex model, Yangian quantum group,
%boson-fermion duality relation

\newpage

\baselineskip=16pt
\no \section{Introduction }
\renewcommand{\theequation}{1.{\arabic{equation}}}
\setcounter{equation}{0}
\medskip

Exactly solvable quantum many particle systems and spin chains with long-range
interactions~\cite{Ca71,Su71,Su72,OP83,Ha88,Sh88,Po93,HW93,Ha96,CS02} have attracted much
attention due to their appearance in apparently diverse branches of 
physics and mathematics like
generalized exclusion statistics~\cite{Ha96,MS94,Po06, BGS09, BH09}, 
quantum Hall effect~\cite{AI94}, quantum
electric transport in mesoscopic systems~\cite{BR94,Ca95}, random matrix theory~\cite{TSA95},
multivariate orthogonal polynomials~\cite{Fo94,UW97,BF97npb} and Yangian quantum
groups~\cite{BGHP93,Hi95npb,BBHS07}. The study of this type of models with long-range
interaction was initiated by Calogero~\cite{Ca71}, who has found  
the exact spectrum of an $N$-particle system on a line with two-body interactions inversely
proportional to the square of their distances and subject to a confining harmonic potential. 
The Hamiltonian of such rational Calogero model
%associated with the $ A_{ N - 1 } $ root system, 
may be written in the form~\cite{Ca71,Su71}
\begin{eqnarray}
 H_{A}=\frac{ 1 }{ 2 } \sum_{ i = 1 }^{N} \big( - \,\frac{\d^2}{\d x_i^2}  
+ \omega^2 x_i^2 \, \big) +
\alpha ( \alpha -  1 ) \sum_{ 1 \leqslant i < j \leqslant N } \frac {1 }
{( x_i - x_j )^2},
\label{a1}
\end{eqnarray}
where $ \alpha ~(> \frac{1}{2} ) $  is a free parameter.  
It has been found that, this Hamiltonian yields a quantum integrable model  
associated with the $ A_{ N - 1 } $ root system and it is possible to
construct generalizations of this Hamiltonian for 
other root systems while preserving the quantum integrability property~\cite{OP83,KPS00,LS04}.  
In particular, for the case of $ D_N $ root system, the Hamiltonian of Calogero
model is given by 
\begin{eqnarray}
H_D = \frac{1}{2} \sum_{i = 1}^{N} \big(-\,\frac{\d^2}{\d x_i^2}
 + \omega^2x_i^2 \, \big)+\nu(\nu-1)
\sum_{ 1 \leqslant i < j \leqslant N }
\left[\frac{1}{(x_i-x_j)^2} + \frac{1}{(x_i+x_j)^2}
\right],
\label{a2}
\end{eqnarray}
where $ \nu ~(> \frac{1}{2} ) $  is a free parameter. Furthermore, the Hamiltonian of 
 Calogero model associated with the $B_N$ root system is related to its 
$D_N$ counterpart as  
\bea
 H_B  = H_D + \frac{1}{2}\, {\rho ( \rho - 1 )} \, \sum_{ i = 1  }^{ N } 
\frac{1}{ x_j^2}  \, ,
\label{a3}
\eea
where $ \rho ~(> \frac{1}{2} )$ is another free parameter corresponding to the one-body
potential.

Due to Eq.(\ref{a3}), one may naively think that the $D_N$ type of Calogero model is just a 
special case of its $B_N$ counterpart and all physically relevant properties
of the former model can be obtained from those of the latter model by simply taking the 
$\rho\rightarrow 0$ limit.  However, the spectra of   
Calogero models associated with all root systems can be calculated  
by acting the corresponding Hamiltonians on Coxeter invariant 
Polynomials \cite{KPS00,LS04}.  It turns out that, contrary to the naive expectation, 
spectrum of the $D_N$ type of Calogero model
can not be reproduced from its $B_N$ counterpart  by taking the 
$\rho\rightarrow 0$ limit. Moreover,
the spectra of  $BC_N$ and $D_N$ types of Calogero models along with their spin generalizations
have been computed recently by finding out appropriate sets of basis vectors
on which the corresponding auxiliary Hamiltonians
and Dunkl operators act as some triangular matrices \cite{BFGR08,BFG09}. 
Again it is found that, 
spectra of these $D_N$ type of models
can not be reproduced from their $B_N$ counterparts  as some special cases. 
Consequently, the $D_N$ type of Calogero model
and its spin generalization should be considered 
as some singular limits of their $B_N$ counterparts.

Even though the eigenvalue problem of the
$D_N$ type of Calogero model (\ref{a2}) has been studied earlier 
through different approaches, the connection of the corresponding 
Hilbert space with that of free quantum harmonic oscillators (QHO) 
has not been explored till now. 
In this context it should be noted that, one can solve the eigenvalue problem 
of $A_{N-1}$ and $B_N$ type of Calogero models 
by using  similarity transformations 
which map  these models to a system of $N$ number of decoupled QHO
(up to some additive constants) \cite{GP99, GP99a, So96, NUW00}. 
However, due to the difference of domains on which
these similarity transformations act as nonsingular operators, 
the spectrum of $A_{N-1}$ 
 type of Calogero model differs significantly from that of the $B_N$ type of Calogero model.
More precisely, up to a constant shift of all energy levels,
the spectrum of the $A_{N-1}$ type of Calogero model coincides with that 
of $N$ number of bosonic QHO, 
which corresponds to completely symmetric wave functions \cite{GP99, NUW00}. 
On the other hand, the spectrum of the  
$B_{N}$ type of Calogero model can be identified with a subset of the 
spectrum of $N$ number of bosonic QHO, which corresponds
to completely symmetric as well as  even parity 
wave functions \cite{GP99a, NUW00}. The orthogonality relations 
for the eigenfunctions of both $A_{N-1}$ and $B_N$ types of Calogero models
have also been established \cite{NUW00}. 
The purpose of the present article is to make a connection between the 
Hilbert space of the $D_N$ type of Calogero model (\ref{a2})
and that of QHO, by applying the method of similarity transformation.

The arrangement of this paper is as follows. In Sec.2, we describe the  
similarity transformation which maps this $D_N$ type of Calogero model
 to a system of $N$ number of decoupled QHO. In this section, 
we also find out the domain on which such similarity transformation
acts as a nonsingular operator.  By using these results, in Sec.3 we
solve the eigenvalue problem  of the  $D_N$ type of Calogero model. 
In particular, we construct the eigenfunctions of this Calogero model  
from those of bosonic harmonic oscillators having 
either all even parity or all odd parity.
We also show that eigenfunctions of this model are orthogonal with respect
to a nontrivial inner product, which has a close connection with the  
quasi-Hermiticity property of the corresponding conserved quantities.
In Sec.4 we make some concluding remarks. 

\no \section{Similarity transformation and its domain }
\renewcommand{\theequation}{2.{\arabic{equation}}}
\setcounter{equation}{0}
\medskip

Our aim is to solve the eigenvalue problem given by 
\beq
H_D \, \psi(x_1, \ldots ,x_N) \, = \,  E \, \psi(x_1, \ldots ,x_N) \, ,
\label{b1}
\eeq
by constructing a similarity transformation which would map $H_D$ (\ref{a2}) to a set of 
decoupled quantum harmonic oscillators (QHO). It is well known that,  
the ground state wave function for the 
$D_N$-type Calogero model can be expressed as 
\begin{eqnarray}
\psi_0(x_1, \ldots ,x_N) = \prod_{ 1 \leqslant i < j \leqslant N }
|x_i^2-x_j^2|^\nu\hspace{.2 cm} e^{-\frac{1}{2}\omega
\sum\limits_{i = 1}^N x_i^2},
\label{b2}
\end{eqnarray}
and the ground state energy is given by                                
\begin{eqnarray}
 E_0 = \frac{1}{2}N\omega + \nu N(N-1)\omega.
\label{b3}
\end{eqnarray}
By using the `operator form' of this ground state wave function, 
we perform a similarity transformation on $H_D$ as
%and obtain $\tilde{H}_D$ as 
\begin{eqnarray}
 \tilde{H}_D & = & \psi_0^{-1}(H_D-E_0)\psi_0  \nn \\
           & = & \sum_{i = 1}^{N}\big(-\frac{1}{2} \frac{\d^2}{\d x_i^2} 
+ \omega \, x_i\frac{\d}{\d x_i} \, \big)-2\nu \sum_{ 1 \leqslant i < j \leqslant N }
\frac{1}{(x_i^2-x_j^2)}\big (x_i \frac{\d}{\d x_i}  -x_j\frac{\d}{\d x_j} \big).  
\label{b4}                
 \end{eqnarray}
Note that 
the eigenvalue equation (\ref{b1}) for $H_D$ can equivalently be expressed as an eigenvalue 
equation for $\tilde{H}_D$:
\beq
\tilde{H}_D \, \phi(x_1, \ldots ,x_N) = (E-E_0) \, \phi(x_1, \ldots ,x_N)  \, ,
\label{b5}
\eeq
where the corresponding eigenfunctions are related as 
\begin{eqnarray}
\psi(x_1, \ldots ,x_N) \,  = \, \psi_0(x_1, \ldots ,x_N) \phi(x_1, \ldots ,x_N) \, .
\label{b6}
\end{eqnarray}

Let us now consider the Euler operator ($ O_E $) and $D_N$ type of Lassalle operator ($ O_L$)
given by 
\begin{eqnarray}
  O_E  =  \sum_{i = 1}^N x_i \frac{\d}{\d x_i}, ~~~~
 O_L  =  \sum_{i = 1}^N  \frac{\d^2}{\d x_i^2} 
+ 4\nu\sum_{ 1 \leqslant i < j \leqslant N }
\frac{1}{(x_i^2-x_j^2)}\big (x_i \frac{\d}{\d x_i}  -x_j\frac{\d}{\d x_j} \big) \, ,
\label{b7}
\end{eqnarray}
which satisfy the commutation relation
\begin{eqnarray}
\left[O_L, O_E\right] = 2  O_L.
\label{b8}
\end{eqnarray}
In terms of these two operators, $\tilde{H}_D$ in Eq.(\ref{b4}) can be written in a  
compact form like   
\begin{eqnarray}
 \tilde{H}_D = \omega  O_E -\frac{1}{2} O_L \, .
\label{b9}
\end{eqnarray} 
By using the commutation relation (\ref{b8}) and the well known Baker-Campbell-Hausdorff (BCH)
formula, we find that  $\tilde{H}_D$ (\ref{b9}) can be 
transformed into the Euler operator as 
\begin{eqnarray}
e^{\frac{1}{4\omega} O_L} \, \tilde{H}_D  \, e^{-\frac{1}{4 \omega} O_L} \, = \,  \omega \, O_E.
\label{b10}
\end{eqnarray}
Let us now define the Laplacian operator as $ \nabla^2 \equiv \sum_{i = 1}^N
 \frac{\d^2}{\d x_i^2}$. It is easy to see that this Laplacian operator and Euler operator
satisfy the commutation relation: $[\nabla^2,  O_E]=2 \nabla^2$. 
By using this commutation relation  and the BCH
formula, one finds that
\begin{eqnarray}
 e^{-\frac{1}{4 \omega} \nabla^2} \left (\omega \, O_E \right)
e^{\frac{1}{4\omega}\nabla^2 } = \omega \, O_E
-\frac{1}{2}\nabla^2 \equiv \bar{H}. 
\label{b11}
\end{eqnarray}
Next, we introduce the operator $ X^2 \equiv \sum_{i = 1}^N x_j^2 $, which 
satisfies the commutation relations
$$
[ O_E, X^2 ]=2 X^2,  ~~~~
[\nabla^2, X^2]=2 (2 \hat O_E + N ) \, .
$$ 
By using these commutation relations  and the BCH
formula, it is easy to find that
\begin{eqnarray}
 e^{-\frac{1}{2}\omega X^2} \hspace{.1cm}\bar{H} \hspace{.1cm}e^{\frac{1}{2}\omega X^2} =
H_{QHO}-\frac{1}{2} N \omega,
\label{b12}
\end{eqnarray}
where
\begin{eqnarray}
H_{QHO} = \frac{1}{2}\sum_{j = 1}^N (- \,\frac{\d^2}{\d x_j^2}+ \omega^2x_j^2) \, ,
\label{b13}
\end{eqnarray}
represents the Hamiltonian of $N$ number of decoupled QHO. Combining the relations 
(\ref{b10}), (\ref{b11}) and (\ref{b12}), we find that 
\begin{eqnarray}
 T^{-1} \, \tilde{H}_D  \, T \, = \, H_{QHO} -\frac{1}{2} N \omega
%\omega\sum_{j=1}^N \hat{n}_j,
\label{b14}
\end{eqnarray}
where 
\beq 
 T \, = \,  e^{-\frac{1}{4\omega}
O_L}\, e^{\frac{1}{4\omega}\nabla^2}\, e^{\frac{1}{2}\omega X^2} \, .
\label{b15}
\eeq

Next, we try to construct the Hilbert space of  Hamiltonian $\tilde{H}_D$
from that of $H_{QHO}$,
by using the similarity transformation (\ref{b14}). 
To this end, we consider the
creation and annihilation operators of QHO given by
\beq
a_j = \frac{ i }{ \sqrt{ 2 \omega }} ( p_j - i \omega x_j ),~~~~
a_j^{\dagger} = \frac{ -i  }{ \sqrt{ 2 \omega }}( p_j + i \omega x_j ), 
\label{b16}
\eeq
where $p_j \equiv -i \frac{\d}{\d x_j}$.
% and $j \in \{1,2, \cdots, N\}$. 
These operators satisfy the standard bosonic commutation relation:
$[a_i, a_j]=0,~ [a_i^\dagger , a_j^\dagger]=0 ,~ [a_i , a_j^\dagger]= \delta_{ij}$, 
for all $i,j \in \{1,2, \cdots, N\}$.
In terms of these creation and annihilation
operators, the number operator for the $ j$-th oscillator is defined as 
\begin{eqnarray}
n_j \equiv a_j^{\dagger} a_j = \frac{1}{2\omega}(p_j^2 + \omega^2 x_j^2)-\frac{1}{2}, 
\label{b17}
\end{eqnarray}
and $H_{QHO}$ in Eq.(\ref{b13}) can be expressed as 
\beq
H_{QHO} = \omega \, \sum\limits_{j=1}^N {n}_j  \, .
\label{b18}
\eeq
 Since the number operators ($n_i$'s) 
are mutually commuting conserved quantities for $H_{QHO}$, corresponding
simultaneous eigenfunctions are given by 
\begin{eqnarray}
 |\lambda_1,\lambda_2, \ldots ,\lambda_N\rangle = 
\prod_{j=1}^N  (a_j^{\dagger})^{\lambda_j}|0\rangle,
\label{b19}
\end{eqnarray}
where $ \lambda_j$ $(\in {\mathbb{Z}}^{\geq 0})$ 
 is the quantum number associated with the number operator $n_j$ and
$ a_j |0 \rangle = 0 $ for all values of $j$. Due to the existence of the 
similarity transformation (\ref{b14}), one may naively think that 
the wave functions defined as 
\beq
\v \phi_{\lambda_1,\lambda_2, \ldots ,\lambda_N} \r \equiv  
T \v\lambda_1,\lambda_2, \ldots ,\lambda_N\rangle \, , 
\label{b20}
\eeq
would be eigenfunctions of $\tilde{H}_D$ with eigenvalue 
$E_{\lambda_1,\lambda_2, \ldots
,\lambda_N} =\omega \, \sum_{j=1}^N {\lambda}_j - \frac{1}{2}N \omega$.
However, before reaching to this conclusion,  it is important to find out 
the domain of the operator $T$ by checking whether $\v \phi_{\lambda_1,\lambda_2, \ldots
,\lambda_N} \r$ represents a nonsingular, square integrable wave function. 
To this end,  we rewrite the operator $T$ in Eq.(\ref{b15}) as 
$T= e^{ - \frac{ 1 }{ 4 \omega }  O_L} \chi$, where  
 $ \chi \equiv e^{\frac{ 1 }{ 4 \omega} \nabla^2}\hspace{.1cm} e^{ \frac{  1}{ 2 } \omega X^2
} $. Through direct calculation it can be shown that $\chi$ satisfies the relations
\begin{eqnarray}
 %\chi \, a_j^{ \dagger } \, \chi^{ -1 } \, = \, \sqrt{2\omega} \,  x_j \, ,
%which leads to 
 \chi \, (a_j^{\dagger})^{\lambda_j} \, =\, (2\omega)^{\frac{\lambda_j}{2}} \, x_j^{\lambda_j} \,
\chi \, ,~~~~ \chi\,  \v 0 \rangle = 1 \, .
\label{b21}
\end{eqnarray}
By using these relations, we find that $\v \phi_{\lambda_1,\lambda_2, \ldots ,\lambda_N} \r$
in Eq.(\ref{b20}) can be expressed as (in the coordinate representation)
\begin{eqnarray}
\v \phi_{\lambda_1,\lambda_2, \ldots ,\lambda_N} \r  
= (2\omega)^{\frac{1}{2}\sum\limits_{j=1}^{N}
\lambda_j} \, e^{ - \frac{ 1 }{ 4 \omega }  O_L}
\big(x_1^{\lambda_1} x_2^{\lambda_2} \ldots
x_N^{\lambda_N}\big) \, .
\label{b22} 
\end{eqnarray}
From the above equation it is evident that, 
$\v \phi_{\lambda_1,\lambda_2, \ldots ,\lambda_N} \r $
would be a singular wave function, if the action of the 
Lassalle operator $ O_L $ on the monomial $ x_1^{\lambda_1} x_2^{\lambda_2} \ldots
x_N^{\lambda_N} $ leads to a singularity. By using Eq.(\ref{b7}), we get   
\begin{eqnarray}
&O_L (x_1^{\lambda_1}\, x_2^{\lambda_2} \ldots x_N^{\lambda_N})
=\sum\limits_{j=1}^N \lambda_j(\lambda_j-1)x_1^{\lambda_1} 
\ldots
x_j^{\lambda_j-2} \ldots x_N^{\lambda_N}  \nn \\ 
&\hskip 6 true cm + \,  4\nu\sum\limits_{1\leq i<j \leq N}\frac{\lambda_i-\lambda_j} {x_i^2-x_j^2}
\, \left(x_1^{\lambda_1}\, x_2^{\lambda_2} \ldots x_N^{\lambda_N}\right). 
\label{b23}
\end{eqnarray}
Note that first term in the r.h.s. of Eq.(\ref{b23})
 seems to be singular at $ x_j=0 $, whenever $ \lambda_j $ takes the value
$0$ or $1$. However, the presence of the coefficient $ \lambda_j(\lambda_j-1) $
within this term precludes that possibility. On the other hand, 
the second term in the r.h.s. of Eq.(\ref{b23})
has pair of simple poles at $ x_i=x_j $ and $ x_i=-x_j $. 
Consequently, successive action of the Lassalle 
operator on the monomial $x_1^{\lambda_1}\, x_2^{\lambda_2} \ldots x_N^{\lambda_N}$ 
yields essential singularities at these points. Due to such singularities,
$\v \phi_{\lambda_1,\lambda_2, \ldots ,\lambda_N} \r $ in Eq.(\ref{b22})
does not represent a square integrable wave function.

As a first step to get rid of the above mentioned singularity,  one may 
apply the similarity transformation $T$ on the completely symmetrized 
number states of QHO, as was done earlier \cite{GP99, GP99a, NUW00}
both in the cases of $A_{N-1}$ type and $B_N$ 
type of Calogero models. For the sake of convenience, we consider a system
consisting of two free harmonic oscillators and 
define the corresponding symmetrized number states as 
\begin{eqnarray}
\v \lambda_1, \lambda_2\r_s \equiv \tau 
\big( \v \lambda_1,\lambda_2\r + \v \lambda_2, \lambda_1\r \big), 
\label{b24}
\end{eqnarray}
where we assume that $\lambda_1 \leq \lambda_2$, and set 
$\tau = 1$ for $\lambda_1 < \lambda_2$ and $\tau = 1/2$ for $\lambda_1 = \lambda_2$.
Applying the similarity transformation $T$ on such symmetrized 
number state and using Eq.(\ref{b21}), we obtain 
\begin{eqnarray}
\v \phi_{\lambda_1,\lambda_2 }^s \r \equiv 
 T \v \lambda_1, \lambda_2 \r_s \,=\,\tau \, (2\omega)^{\frac{1}{2}(\lambda_1+\lambda_2)} \,
 e^{-\frac{1}{4\omega} O_L}
\, (x_1^{\lambda_1}  x_2^{\lambda_2} + x_1^{\lambda_2} x_2^{\lambda_1}). 
\label{b25}
\end{eqnarray}
By using Eq.(\ref{b7}), one finds that
\begin{eqnarray}
 &O_L (x_1^{\lambda_1}  x_2^{\lambda_2} + x_1^{\lambda_2} x_2^{\lambda_1}) 
\, = \,  \lambda_1(\lambda_1-1)(x_1^{\lambda_1-2}x_2^{\lambda_2} +
x_1^{\lambda_2} x_2^{\lambda_1-2}) \hskip 4 cm \nn \\ 
 & \hskip .4 cm + \, \lambda_2(\lambda_2-1) 
(x_1^{\lambda_1}  x_2^{\lambda_2-2} + x_1^{\lambda_2-2} x_2^{\lambda_1}) %\\ & + &
+ 4\nu(\lambda_2-\lambda_1)x_1^{\lambda_1}x_2^{\lambda_1}
\left(\frac{x_1^{\lambda_2-\lambda_1}-x_2^{\lambda_2-\lambda_1}}{
x_1^2-x_2^2 } \right).
\label{b26}
\end{eqnarray}
Note that the singularities in the r.h.s. of the above
equation can be removed completely, if we
restrict the value of $ \lambda_2 -\lambda_1$ 
to be an even integer. 
Indeed, by setting $\lambda_2 -\lambda_1 = 2 m $,
 where $ m \in { \mathbb{ Z }}^{ \geq 0 }, $ and defining symmetrized
polynomials like 
\begin{eqnarray*}
 \varphi_{\lambda_1,\lambda_2}\, = \, \tau  \left(x_1^{\lambda_1}  x_2^{\lambda_2} + x_1^{\lambda_2} x_2^{\lambda_1}\right) \, ,
\end{eqnarray*}
one can express Eq.(\ref{b26})  in the form
\begin{eqnarray*}
  O_L \, \varphi_{\lambda_1,\lambda_2 } 
= \lambda_2(\lambda_2-1) C_{\lambda_1,\lambda_2} \,
\varphi_{\lambda_1, \lambda_2-2} +
\lambda_1(\lambda_1-1) \varphi_{\lambda_1-2, \lambda_2} 
+ 8\nu m \sum_{ i= 1}^t 
\varphi_{\lambda_1+2i-2, \lambda_2-2i} \, ,
\end{eqnarray*}
where $ C_{\lambda_1,\lambda_2} = (1 - \delta _{\lambda_1, \lambda_2} + 
\delta_{\lambda_1, \lambda_2-2})$ 
and  $t=[(m + 1 )/2] $, with $ [ x ] $
denoting the integer part of $ x $. From the r.h.s. of the above 
equation it is clear that, repeated actions of 
$O_L$ on $\varphi_{\lambda_1,\lambda_2 }$ do not produce 
any singularity. Consequently, $\v \phi_{\lambda_1,\lambda_2 }^s \r$ in Eq.(\ref{b25})
would represent a nonsingular and square integrable eigenfunction 
of $\tilde{H}_D$, provided $ \lambda_2 -\lambda_1$ 
is taken as an even integer. 

In analogy with the two particle case, 
as considered in Eq.(\ref{b25}), one can construct completely symmetrized
 states like $\v \phi_{\lambda_1,\lambda_2, \cdots ,\lambda_N }^s \r$ for the 
general $N$ particle case.  
Such construction will be discussed in the next section. 
Proceeding in a similar way as has been done earlier  
in the case of $B_N$ model \cite{NUW00}, it can be shown that 
$\v \phi_{\lambda_1,\lambda_2, \cdots ,\lambda_N }^s \r $ would represent
a nonsingular eigenfunction of $\tilde{H}_D$,
provided $ \lambda_j -\lambda_i$ are 
even integers for all $i,j \in \{1,2,\cdots , N \}$. Note that the above  
condition is satisfied if either all $\lambda_i$'s are even integers,
i.e. of even parity, or all $\lambda_i$'s are odd integers,
i.e. of odd parity. Therefore, the Hilbert space  of $D_N$ type 
of Calogero model (denoted by $\mathcal H$) can be decomposed as 
\beq
\mathcal {H} =\mathcal {H}_0 \oplus \mathcal {H}_1 \, ,
\label{b27}
\eeq
where the subspace $\mathcal {H}_0$ is made of states with even parity and 
the subspace $\mathcal {H}_1$ is made of states with odd parity.
It should be noted that,
the decomposition of the Hilbert space given in Eq.(\ref{b27})
and corresponding eigenvalues of $H_D$ (see Eq.(\ref{c10}))
 was found earlier through a completely different approach involving
the $D_N$ type of Dunkl operators \cite{BFG09}. 
However, the present approach through similarity
transformation not only enables us to reproduce these 
results, but also leads to explicit expressions 
for the corresponding eigenfunctions in a simple way. 

As we have mentioned earlier that, there exists a
similarity transformation which maps the $B_N$ 
type of Calogero model (\ref{a3}) to a system of decoupled QHO \cite{NUW00}.
At $\rho \rightarrow 0$ limit, that similarity transformation formally reduces to the 
the presently considered similarity transformation $T$ (\ref{b15}). 
However, it is important to observe that, the domains of these two similarity 
transformations do not match with each other. 
To verify this thing, we note that the   
 Lassalle operator $  O_L^{ ( B) } $ associated with the 
 $ B_N $ type of Calogero model (\ref{a3}) is given by \cite{NUW00}
\begin{eqnarray*}
 O_L^{( B )} = \sum_{i = 1}^N \big( \frac{\d^2}{\d x_i^2}
 + 2 \rho \, \frac{1}{x_i} \frac{\d}{\d x_i} \big) \hspace{.1cm}
+ 4 \nu \sum_{ 1 \leqslant i < j \leqslant N }
\frac{ 1 }{ ( x_i^2 - x_j^2 )}
 \big( x_i \frac{\d}{\d x_i} - x_j \frac{\d}{\d x_j} \big) \, .
\end{eqnarray*}
Action of this $  O_L^{( B )} $ on the monomial 
$x_1^{\lambda_1} x_2^{\lambda_2} \ldots x_N^{\lambda_N}$ yields
\begin{eqnarray}
 O_L^{( B )} (x_1^{\lambda_1} x_2^{\lambda_2} \ldots x_N^{\lambda_N}) 
= \sum_{ j = 1 }^N \left\{ \lambda_j(\lambda_j-1) +  2 \rho
\lambda_j \right\} x_1^{\lambda_1} \ldots x_j^{\lambda_j-2} \ldots 
x_N^{\lambda_N} \nonumber \\ + 
4 \nu \sum_{ 1 \leqslant i < j \leqslant N }\frac{ \lambda_i - \lambda_j } { x_i^2 -
x_j^2 }\hspace{.1cm} (x_1^{\lambda_1} x_2^{\lambda_2} \ldots
x_N^{\lambda_N}). 
\label{b28}
\end{eqnarray}
Comparing the first terms in the r.h.s. of Eqs.~(\ref{b23}) and (\ref{b28}),  we find that the
coefficient $ \lambda_j(\lambda_j-1) $ in the former equation is
replaced by the coefficient $ \lambda_j(\lambda_j-1)+2 \rho \lambda_j $ in the latter equation.
 Consequently, unlike the case of $D_N$ type of  
Calogero model, the first term in the r.h.s. of 
Eq.(\ref{b28}) picks up a singularity at $ x_j = 0 $ for the choice $\lambda_j=1$. 
Moreover,  
successive action of $O_L^{( B )}$ yields this type of singularity
at $ x_j =  0 $ for any odd value of $ \lambda_j $. 
%Note that this type of 
%singularities can not be eliminated by acting $O_L^{( B )}$ on the 
%completely symmetrized form of the 
%monomial  $x_1^{\lambda_1} x_2^{\lambda_2} \ldots x_N^{\lambda_N}$. 
Thus the similarity transformation associated with the $B_N$ 
type of Calogero model generates singularity while acting on the
completely symmetric states of QHO with odd parity.
On the other hand, all singularities appearing in Eq.(\ref{b28})
can be eliminated by acting $O_L^{( B )}$ on the 
completely symmetrized form of the 
monomial  $x_1^{\lambda_1} x_2^{\lambda_2} \ldots x_N^{\lambda_N}$ and restricting
all $\lambda_i$'s to be even integers \cite{NUW00}.  
%Since the similarity transformation associated with the $B_N$ 
%type of Calogero model does not generate any singularity only if it 
%acts on the completely symmetric states of QHO with even parity,  
Consequently, the Hilbert space of this $B_N$ type of Calogero model 
can be constructed by using such completely symmetric states with even parity only.

%%%%%%%%%%%%%%%%%%%%%%%%%%%%%%%%%%%%%%%%%%%%%%%%%%%%%%%%%%%%%%%%%%%%%%%%%%%%%%%%%%%%%%%%%%%%%%%%%%%%%
%%%%%%%%%%%%%%%%%%%%%%%%%%%%%%%%%%%%
\no \section{Construction of Eigenfunctions }
\renewcommand{\theequation}{3.{\arabic{equation}}}
\setcounter{equation}{0}
\medskip

Here our aim is to construct the eigenfunctions of  
the Hamiltonian $H_D$ (\ref{a2}) for the general $N$ particle case
%from those of $H_{QHO}$,
%by applying the similarity transformation (\ref{b14}).
and find out the scalar product of such eigenfunctions.
To this end, we consider a set of nonnegative integers like $
{\vec \lambda}\equiv \{\lambda_1, \lambda_2, \ldots,
\lambda_N \} $,  subject to
restriction that all \hspace{.05cm}$ \lambda_i $'s 
have either positive parity 
or negative parity and the ordering  
\hspace{.2cm}$ \lambda_1 \geq \lambda_2 \geq \ldots
\geq
\lambda_N \geq 0. $
%By following the approach of Ref.\cite {NUW00} in the case of $B_N$ Calogero model,
One can construct a symmetrized number state associated with ${\vec \lambda}$ as
\begin{eqnarray}
|{\vec \lambda} \rangle_s \equiv \sum_{ \sigma \in S_N }|\lambda_{ \sigma_1}, \ldots , \lambda_{
\sigma_N }
\rangle = \varphi_{\vec \lambda}( {\bf{a}}^{ \dagger }
)| 0 \rangle,
\label{c1}
\end{eqnarray}
where $ \varphi_{\vec \lambda}( {\bf{x}} ) $ is a 
completely symmetric function of ${\bf x}$ defined by \cite{NUW00}
\begin{eqnarray}
\varphi_{\vec \lambda}( {\bf{x}} ) = \sum_{ \sigma \in S_N } x_1^{ \lambda_{ \sigma_1 }} x_2^{
\lambda_{
\sigma_2 }} \ldots x_N^{ \lambda_{ \sigma_N }} \, ,
\label{c2}
\end{eqnarray}
with \hspace{.02cm} $ {\bf{x}} \equiv \{ x_1, x_2, \ldots, x_N \}
 \in {\mathbb {R}}^N $, and the summation runs over distinct permutations
 so that each monomial appears only once. 
It may be noted that, 
for the particular case $N=2$, $|{\vec \lambda} \rangle_s $ in Eq.(\ref{c1}) reproduces $\v
\lambda_1, \lambda_2\r_s$ in Eq.(\ref{b24}). 

Next, by using the number operators, we define 
a set of mutually commuting Hermitian operators like  
\begin{eqnarray}
P_l({\bf n}) \equiv \sum_{j=1}^N n_j^l \, , 
\label{c3}
\end{eqnarray}
where $l \in \{1,2, \cdots , N\}$. Due to Eq.(\ref {b18}), 
it follows that $H_{QHO}=\omega P_1({\bf n})$. 
Hence $P_l({\bf n})$'s 
represent a complete set of mutually commuting conserved quantities for the QHO.
It is evident that the symmetrized number states (\ref{c1})
are simultaneous eigenfunctions of these conserved quantities:
\begin{eqnarray}
 P_l( {\bf{n}} ) \, | \vec \lambda \rangle_s \, = 
\, P_l(\vec \lambda) \, | \vec \lambda \rangle_s \, , 
\label{c4}
\end{eqnarray}
 where $P_l(\vec \lambda) =  \sum_{j=1}^N \lambda_j^l \, .$ 
%Since the states (\ref{c1}) are non-degenerate eigenfunctions of the Hermitian operators
%(\ref{c3}), they constitute a complete set of orthogonal bases for the QHO.
We define the dual bases for the states (\ref{c1})  as
\begin{eqnarray}
\langle \vec \lambda |_s \equiv \langle 0 | 
\hspace{.1cm} \varphi_{\vec\lambda} ( {\bf{a}^\dagger} ),
\label{c5} 
\end{eqnarray}
where $\l 0|$ is defined through the
relations \hspace{.1cm}$ \langle 0 |\hspace{.05cm} a_j^\dagger = 0, $ for all values 
of $j$. By using the bosonic commutation relations satisfied by the creation and annihilation
operators, the orthogonality relations among the 
scalar products of the symmetrized number states may be obtained as 
\begin{eqnarray}
\langle \mu |\lambda \rangle_s \, =\,
\delta_{ \vec {\lambda}, \, \vec{\mu} } \,
 \l 0 | 0 \r \, N! \,
\prod\limits_{ j = 1 }^r \frac{( l_j ! )^{ k_j }}{k_j!}
  \, ,
% \frac{ N ! }{ \prod\limits_{ i = 1 }^r k_i ! } \, \prod\limits_{j = 1}^r
%( l_j! )^{ k_j } \, \langle 0 | 0 \rangle \,  \delta_{\vec \lambda, \, \vec \mu} ~ ,
\label{c6}
\end{eqnarray}
where the notation
$ \langle\vec \mu |_s \ldotp |\vec \lambda 
\rangle_s \equiv \langle  \mu |  \lambda \rangle_s \, $ is used, 
$\vec \lambda$ is written in the form
 \begin{eqnarray}
\vec \lambda  = 
 \{\, \overbrace{ l_1, l_1, \ldots , l_1 }^{ k_1 } ,\hspace{.1cm}\overbrace{ l_2, l_2,
\ldots , l_2 }^{ k_2 },\hspace{.1cm}
\ldots , \overbrace{ l_r, l_r, \ldots , l_r }^{ k_r } \, \}  \, ,
%\\        & = & \{\lambda_1, \ldots , \lambda_N \},
\label{c6a}  
\end{eqnarray}
such that $ \sum\limits_{ i = 1 }^r k_i = N,  $ and  
$\l 0 | 0\r=\left( \int_{-\infty}^{\infty} e^{-\omega x^2} dx \right)^N 
=\left( \frac{\pi}{\omega} \right)^\frac{N}{2}$.

By applying the operator $T$ (\ref{b15}) on the 
symmetrized number state 
$ |{\vec \lambda} \rangle_s $ (\ref{c1}),
and using the relations (\ref{b14}) and (\ref{b21}), 
we obtain the eigenfunctions for $\tilde{H}_D$ as
\begin{eqnarray}
\v \phi_{\vec \lambda }^s  \r\equiv 
 T |{\vec \lambda} \rangle_s \,=\, (2\omega)^{\frac{1}{2}\sum\limits_{i=1}^N\lambda_i} \,
 e^{-\frac{1}{4\omega} O_L}\varphi_{\vec \lambda}( {\bf{x}} ) \, ,
\label{c7}
\end{eqnarray}
with eigenvalues given by 
\beq
{\tilde E}_{\lambda_1,\lambda_2, \ldots ,
\lambda_N} =\omega \, \sum_{j=1}^N {\lambda}_j - \frac{1}{2}N \omega \, .
\label{c8}
\eeq 
Proceeding in a similar way as has been done earlier 
in the case of $B_N$ model \cite{NUW00}, it can be shown that 
$\v \phi_{\vec \lambda }^s \r$ in Eq.(\ref{c7}) represents    
%a complete set of 
nonsingular and square integrable eigenfunctions 
for $\tilde{H}_D$. Let us now define an operator $\mathcal{T}$ as 
\beq
{\mathcal T} \equiv \psi_0(\mathbf x) T =\psi_0(\mathbf x)
 e^{-\frac{1}{4\omega}
O_L}\, e^{\frac{1}{4\omega}\nabla^2}\, e^{\frac{1}{2}\omega X^2} \, , 
\label{c8a}
\eeq 
where $\psi_0(\mathbf x)$ is the `operator form' of the ground state 
wave function (\ref{b2}). By using Eqs. (\ref{b6}) and (\ref{c7}), 
we obtain the eigenfunctions for the original Calogero Hamiltonian 
${H}_D$ (\ref{a2}) as
\beq
\v \psi_{\vec \lambda }^s \r =  \mathcal{T}|{\vec \lambda} \rangle_s
%= \psi_0(\mathbf x) \phi_{\vec \lambda }^s 
\, =\, (2\omega)^{\frac{1}{2}\sum\limits_{i=1}^N\lambda_i} \,\psi_0(\mathbf x) \, 
 e^{-\frac{1}{4\omega} O_L}\varphi_{\vec \lambda}( {\bf{x}} ) \, .
\label{c9}
\eeq
Subsequently, by using Eq.(\ref{b5}), we obtain the corresponding eigenvalues as 
\beq
 E_{\lambda_1,\lambda_2, \ldots ,
\lambda_N} = 
{\tilde E}_{\lambda_1,\lambda_2, \ldots ,
\lambda_N} + E_0 =\omega \, \sum_{j=1}^N {\lambda}_j +\nu N (N-1) \omega \, ,
\label{c10}
\eeq 
where all \hspace{.05cm}$ \lambda_j $'s 
have the same parity and  they are ordered as   
\hspace{.2cm}$ \lambda_1 \geq \lambda_2 \geq \ldots \lambda_N \geq 0$.

In this context it should be noted that, by acting the Hamiltonian of 
the $D_N$ type of Calogero model on the corresponding  Coxeter invariant 
Polynomials, one can get the spectrum of this model in the form \cite{KPS00,LS04} 
\beq
E_{ m_1, \ldots, m_N } = \omega\sum_{ j = 1 }^N m_j f_j + \nu N ( N - 1
) \omega \, ,
\label{c10a}
\eeq
where $f_j = 2j$ for  $j \in \{ 1, \ldots , N - 1 \}$, $f_N = N$ and 
$m_j$'s are arbitrary non-negative integers. To make a connection 
between the eigenvalue relations (\ref{c10}) and (\ref{c10a}), 
we define a mapping between the related quantum numbers as
\beq
\lambda_j = 2 \sum\limits_{ i = j }^{ N - 1 } m_i + m_N.
\label{c10b}
\eeq
Note that this is a one-to-one mapping, whose inverse is given by 
$n_j = \frac{1}{2}( \lambda_j - \lambda_{ j + 1 } ) \, $
for $ j \in \{ 1, \ldots , N - 1 \}$ and $n_N = \lambda_N$. 
Substituting Eq.(\ref{c10b}) in Eq.(\ref{c10}), and interchanging 
the summations over $i$ and $j$ indices, we find that the spectra
generated by Eq.(\ref{c10}) and Eq.(\ref{c10a}) match exactly.
It is interesting to note that, due to Eq.(\ref{c10b}), 
the parity of $m_N$  determines the parity of all the $\lambda_j$'s. 
Consequently,  the eigenvalues in Eq.(\ref{c10a}) with even (odd) values of $m_N$ 
are associated with the eigenfunctions (\ref{c9})
% of  all even (odd) excitations 
corresponding to the subspace $\mathcal{H}_0 ~(\mathcal{H}_1)$.  
%in Eq.(\ref{b27}). 
 
Let us now define a new set of `creation' and `annihilation' operators 
associated with the original Calogero Hamiltonian ${H}_D$ (\ref{a2}) as
\beq
 b_j^{\dagger} = {\mathcal T} \, a_j^{\dagger} \, {\mathcal T}^{-1} \, , ~~
 \tilde{b}_j = {\mathcal T} \, a_j \, {\mathcal T}^{-1} \, ,
\label{c11}
\eeq
%
%$ \eta_j = b_j^{\dagger}\hspace{.1cm}b_j, $
where $j\in \{1,2, \cdots ,N\}$.  Similar to the case of QHO, these creation and annihilation
operators also satisfy the standard bosonic commutation relation:
\beq
[\tilde{b}_i, \tilde{b}_j]=0,
~~ [b_i^\dagger , b_j^\dagger]=0 ,~~ [\tilde{b}_i , b_j^\dagger]= \delta_{ij} ,
\label{c11a}
\eeq 
for all $i,j \in \{1,2, \cdots, N\}$. However, it should be noted that the operator
 $\mathcal T$ defined in Eq.(\ref{c8a}) is not an unitary operator. Consequently,   
$b_j^\dagger$ is no longer the adjoint operator of $\tilde{b}_j$. 
The vacuum state associated with this new type of creation and annihilation
operators may be defined as 
\beq
\v 0\r_D \equiv {\mathcal T} \v 0\r \, , 
\label{c12}
\eeq
which satisfies the relations $ \tilde{b}_j \v 0\r_D = 0$ for all $j$,  and 
coincides with the ground state wave function (\ref{b2}) of the $D_N$ type of
Calogero model in the coordinate representation. 
Due to such coincidence, the normalization condition for ground state wave function 
of the $D_N$ type of Calogero model\cite{NUW00,BF97, Di97} leads to a 
relation like
\begin{eqnarray}
\l 0 |  {\mathcal T}^\dagger  {\mathcal T} |0 \r  
=  \frac{1}{\omega^{ N \{ \frac{ 1 }{ 2 }+  ( N - 1 ) \nu \} }} 
\prod_{ j = 1 }^N
\frac{ \Gamma ( 1 + j \nu ) \Gamma ( \frac{ 1 }{ 2 } + ( j - 1 ) \nu ) }{ \Gamma ( 1 + \nu ) } \, ,
\label{c13}
\end{eqnarray}
%\begin{eqnarray}
%\langle 0 | 0 \rangle_D = \left( \frac{ 1 }{ \omega } \right)^{ N [ \frac{ 1 }{ 2 } 
%+ \nu ( N - 1
%)]}
%\hspace{.1cm} \hspace{.1cm} \prod_{ j < k }^N
%\frac{ \Gamma ( \nu ( k - j + 1 )) \Gamma( 1 + \nu ( k - j + 1 ))}
%{ \Gamma ( \nu ( k - j )) \Gamma (
%1 + \nu ( k - j ))} \nonumber \\
%\times \prod_{ j = 1 }^N \Gamma ( 1 + \nu ( N - j )) \Gamma ( \frac{ 1 }{ 2 } + \nu ( N - j )),
%\label{c13}
%\end{eqnarray}
where $ \Gamma(z) $ denotes the usual gamma function. The above equation clearly
shows that $\mathcal T$ can not be an unitary operator. As a result, one has to be more 
careful for defining the dual vector corresponding to $\v 0\r_D$. Indeed, by following 
the usual convention, if such dual vector is defined as $\l 0 \v_D= \l 0 \v   
{\mathcal T}^\dagger $, then this dual vector would not be annihilated by the 
left action of the creation operators like $b_j^\dagger$. To bypass this problem, 
we define the dual vector corresponding to  $\v 0\r_D$ in Eq.(\ref{c12}) as 
\beq
\l 0 \v_D \equiv \l 0 \v   
{\mathcal T}^{-1} \, , 
\label{c13a}
\eeq
which satisfies the desired relations $ \l 0 \v_D \, b_j^\dagger = 0$ for all $j$. 
This type of  dual vectors, defined in a rather unconventional way,  
 will be used shortly to construct a nontrivial
inner product in the Hilbert space of the $D_N$ type of Calogero model.  

Applying the relations (\ref{b4}), 
(\ref{b14}) and (\ref{b18}), we can express
the Calogero Hamiltonian ${H}_D$ (\ref{a2}) through the number operators
 associated with $\tilde{b}_j$ and $b_j^\dagger$ as
\begin{eqnarray}
H_D = \omega \sum_{j = 1}^N  \eta_j + \nu N (N-1) \omega \, ,
\label{c14}
\end{eqnarray}
where $\eta_j= b_j^\dagger \, \tilde{b}_j $.  Furthermore, by using Eqs. (\ref{c1}), 
(\ref{c11}) and (\ref{c12}),  it is possible to rewrite the eigenfunctions (\ref{c9}) 
of the $D_N$ type of Calogero model
through symmetric combination of different powers of $b_j^\dagger$'s as
\begin{eqnarray}
\v \psi_{\vec \lambda }^s \r =  \mathcal{T}
\varphi_{\vec \lambda}( {\bf{a}}^{ \dagger }) \mathcal{T}^{-1} \cdot \mathcal{T} | 0 \r
= \varphi_{\vec \lambda}( {\bf{b}}^{ \dagger }) \v 0\r_D  \, .
\label{c15}
\end{eqnarray}
Let us now define the dual vector corresponding to $\v \psi_{\vec \mu }^s \r$ as  
\beq
\l \psi_{\vec \mu }^s \v_D \equiv \l 0 \v_D \, \varphi_{\vec \mu}( {\bf{\tilde{b}}}) \, ,
\label{c15a}
\eeq
which leads to a new inner product between the states $\v \psi_{\vec \lambda }^s \r$ 
and $\v \psi_{\vec \mu }^s \r \,$: 
\begin{eqnarray}
\l\l \psi_{\vec \mu }^s \v \psi_{\vec \lambda }^s \r\r 
\equiv 
%\l \psi_{\vec \mu }^s \v_D \cdot \v \psi_{\vec \lambda }^s \r=
 \l 0 \v_D \, \varphi_{\vec \mu}(
{\bf{\tilde{b}}})   \varphi_{\vec \lambda}( {\bf{b}}^{ \dagger }) \v 0\r_D  \, .
\label{c16}
\end{eqnarray}
Since $b_j^\dagger$ is not the adjoint operator of $\tilde{b}_j$, and $\l 0 \v_D$ 
is not the dual of $\v 0 \r_D$ in the conventional sense, it is obvious that   
the inner product given in the above equation 
is different from the conventional Hermitian inner product.
Furthermore, it should be noted that, the inner product (\ref{c16})
is also different in nature from the inner products used earlier \cite{NUW00} for
the cases of $A_{N-1}$ and $B_N$ types of Calogero models, where 
the duals of the vacuum states were defined in the conventional sense.
Using the bosonic commutation relations (\ref{c11a})  
and expressing $\vec\lambda$ in the form (\ref{c6a}), we find that the  
inner product (\ref{c16}) can be computed as 
\begin{eqnarray}
\l \l \psi_{\vec \mu }^s \v \psi_{\vec \lambda }^s \r \r  = 
\delta_{ \vec {\lambda}, \, \vec{\mu} } \,
N ! \, \l 0 | 0 \r_D \, % \delta_{ \vec {\lambda}, \, \vec{\mu} }
\prod\limits_{ j = 1 }^r \frac{( l_j ! )^{ k_j }}{k_j!}
  \, ,
\label{c17}
\end{eqnarray}
where $\l 0 | 0 \r_D =\langle 0 | 0 \rangle= 
\left( \frac{\pi}{\omega} \right)^\frac{N}{2}$.
Thus  the eigenfunctions (\ref{c15}) of $D_N$ type of Calogero model   
are orthogonal to each other with respect to the inner product (\ref{c16}).

Let us now investigate whether there exists any 
deeper reason for the existence of nontrivial inner product (\ref{c16}), 
which makes the eigenfunctions (\ref{c15}) orthogonal.  
In the following, it will be shown that the integrable structure 
of $D_N$ type of Calogero model plays a crucial role in this matter. 
To this end, we apply a similarity transformation  on the symmetrized conserved
quantities (\ref{c3}) of the QHO and construct a set of mutually commuting 
conserved quantities for the $D_N$ type of Calogero Hamiltonian ${H}_D$ (\ref{a2}) as
\begin{eqnarray}
P_l ( { {\bf{\eta}} } ) \, = \, {\mathcal T} P_l({\bf n})  {\mathcal T}^{-1}
\, =  \, \sum_{j=1}^N ({\eta_j})^l,
\label{c18}
\end{eqnarray}
where $l \in \{ 1, 2, \cdots , N\}$. Since the 
exponential of the Lassalle operator has entered in the definition
of $\mathcal{T}$ in Eq.(\ref{c8a}), 
%unlike the case of usual conserved quantities of a dynamical system, 
$P_l ( { {\bf{\eta}} } )$'s  can not 
be expressed in general as some finite power series  of the canonical variables. 
Moreover, due to nonunitarity of the operator  $\mathcal{T}$,
it follows from Eq.(\ref{c18}) that $P_l ( { {\bf{\eta}} } )$'s
are not Hermitian operators in general with respect to the 
conventional inner product. However,
acting  the operator $\mathcal{T}$
on both sides of Eq.(\ref{c4}), we obtain the relation
\begin{eqnarray}
 P_l ( { {\bf{\eta}} } )\, \v \psi_{\vec \lambda }^s \r  \, = 
\,\Big( \sum_{j=1}^N \lambda_j^l \Big) \, \v \psi_{\vec \lambda }^s \r \, , 
\label{c19}
\end{eqnarray}
which shows that the eigenfunctions (\ref{c15}) 
simultaneously diagonalize all of these mutually commuting conserved operators
with  a set of completely real eigenvalues. 

%To fit these observations in a proper mathematical framework,
In this context, 
it is useful to notice that a set of quasi-Hermitian operators (denoted 
by $A_l$'s) are defined through the relations \cite{SGH92} 
\beq
A_l^\dagger \, = \, \varTheta \, A_l \, \varTheta^{-1} \, , 
\label{c20}
\eeq 
where $\varTheta$ is a Hermitian, positive definite operator.
Combining the operator $\varTheta$ and  
standard inner product $\l \phi \v \psi \r$, one can define
a new inner product as 
\beq 
\l \phi \v \psi \r_{\varTheta} \, 
\equiv \, \l \phi \v \varTheta \, \psi \r  \, ,
\label{c21}
\eeq
where $\v \phi \r$ and $\v \psi \r$ are two arbitrary state 
vectors in the corresponding Hilbert space. It is well known
that, quasi-Hermitian operators satisfying the relations
(\ref{c20}) become Hermitian with respect to the new inner product
defined through Eq.(\ref{c21}). Consequently, quasi-Hermitian operators
yield completely real spectra and the corresponding eigenfunctions
become orthogonal with respect to the inner product
$\l \phi \v \psi \r_{\varTheta}$ given in Eq.(\ref{c21}).
Such quasi-Hermitian operators have been studied recently 
due to their appearance in some parity and time reversal
invariant quantum systems which yield real spectra \cite{BB98, M08, NDD05, SG06}. 

Interestingly, by using Eq.(\ref{c18}), we find that the adjoint of the operators
$P_l ( { {\bf{\eta}} } )$'s can be expressed in the form (\ref{c20}) with 
$\varTheta$ given by
\beq
\varTheta = \left( \mathcal{T} \mathcal{T}^\dagger \right)^{-1}\, .
\label{c22}
\eeq
Hence all $P_l ( { {\bf{\eta}} } )$'s are quasi-Hermitian operators
which, due to Eq.(\ref{c22}),  can be transformed into Hermitian operators
by defining an inner product like  
\beq 
\l \phi \v \psi \r_{\varTheta} \, 
\equiv \, \l \phi \v 
\left( \mathcal{T} \mathcal{T}^\dagger \right)^{-1} \, \psi \r  \, .
\label{c23}
\eeq
Choosing $\v \psi \r= \v \varphi_{\vec \lambda }^s \r,$
$\v \phi \r= \v \varphi_{\vec \mu }^s \r$ and using the above definition 
of the inner product, we obtain
\beq
\l \psi_{\vec \mu }^s \v \psi_{\vec \lambda }^s \r_{\varTheta}
= \l 0 \v \mathcal{T}^\dagger \, \varphi_{\vec \mu}
\left({\bf{b}}\right)
 \left( \mathcal{T} \mathcal{T}^\dagger \right)^{-1}
   \varphi_{\vec \lambda}( {\bf{b}}^{ \dagger }) \v 0\r_D  \, ,
\label{c24}
\eeq
where $ \bf{b} \equiv ({\bf{b}}^\dagger)^\dagger $.  
%\equiv \{ (b_1^\dagger)^\dagger,
%(b_2^\dagger)^\dagger, \cdots , (b_N^\dagger)^\dagger \}$ and 
%$(b_i^\dagger)^\dagger$ is the conjugate of $b_i^\dagger$.
Due to Eq.(\ref{c11}), it follows that
\[
\varphi_{\vec \mu}\left({\bf{b}}\right)
\, = \, \left( \mathcal{T} \mathcal{T}^\dagger \right)^{-1}
\varphi_{\vec \mu}(\bf{\tilde{b}})  
\left( \mathcal{T} \mathcal{T}^\dagger \right)  .
\]
Substituting the above expression to the r.h.s.
of Eq.(\ref{c24}),  we find that this r.h.s.
exactly matches with the r.h.s. of Eq.(\ref{c16}).
Consequently, we get the remarkable relation 
\beq
\l \psi_{\vec \mu }^s \v \psi_{\vec \lambda }^s \r_{\varTheta}
=\l\l \psi_{\vec \mu }^s \v \psi_{\vec \lambda }^s \r\r  \, .
\label{c25}
\eeq
This relation clearly shows that the inner product
 $\l\l \psi_{\vec \mu }^s \v \psi_{\vec \lambda }^s \r\r$ defined in Eq.(\ref{c16}) 
emerges in a natural way from the Hermiticity condition of $P_l(\eta)$'s
given in Eq.(\ref{c18}), 
which are quasi-Hermitian operators
with respect to the conventional inner product.

\no \section{Concluding remarks}
\renewcommand{\theequation}{4.{\arabic{equation}}}
\setcounter{equation}{0}
\medskip

Here we solve the eigenvalue problem of the $D_N$ type of Calogero model (\ref{a2}),
by mapping it to $N$ number of decoupled quantum harmonic oscillators (QHO)
through a similarity transformation. 
Though this similarity transformation apparently looks like  
a special case of the similarity transformation which maps the $B_N$ 
type of Calogero model (\ref{a3}) to a system of decoupled QHO,  
interestingly we find that the domains of these two similarity 
transformations do not match with each other. 
%Consequently,  the spectrum of the $D_N$ type of Calogero model 
%differs significantly from that of its $B_N$ counterpart.
%Up to a constant shift of all energy levels, the spectrum of the  
%$D_{N}$ type of Calogero model coincides with a subset of the 
%spectrum of $N$ number of bosonic harmonic oscillators, which corresponds
%to either all even parity or all odd parity eigenfunctions. 
Applying the similarity transformation operator on
either all even parity or all odd parity eigenfunctions 
of the bosonic QHO, we explicitly construct
the eigenfunctions for the $D_{N}$ type of Calogero model.

It turns out that 
these eigenfunctions for the $D_{N}$ type of Calogero model
 are not orthogonal with respect to the conventional
inner product. However, we find that their orthogonality can be established 
by defining a nontrivial inner product.  To explore some deeper 
reason for the existence of such inner product, we 
again use the method of similarity transformation to
construct a set of mutually commuting conserved quantities for the $D_{N}$ type
 of Calogero model. Even though these conserved quantities are 
quasi-Hermitian operators with respect to the conventional
inner product, they can be transformed to Hermitian operators by using the 
nontrivial inner product which we have mentioned above. Thus the integrable structure
of the $D_{N}$ type of Calogero model plays an important role 
in determining the inner product for which the corresponding
eigenfunctions are orthogonal. In future, we hope 
to explore whether there exists any connection between the presently derived
 conserved quantities for the $D_{N}$ type of Calogero model 
and the conserved quantities for this model 
obtained through the Lax operator approach. Moreover, the 
relation between the $D_N$ type of Jack polynomials and the eigenfunctions
for the $D_{N}$ type of Calogero model obtained through
similarity transformation may also be another interesting topic 
for further investigation.
 
\vskip .20 cm

\noindent {\bf Acknowledgements:} The authors would like to thank Prof. R. Sasaki
for many helpful discussions.

\end{document}